\documentclass{article}
\usepackage{spconf,amsmath,graphicx,cite}
\usepackage{amsfonts}
\usepackage{booktabs}
\usepackage{url}
\usepackage{multicol}
\usepackage{tikz}

\newcommand*{\ditto}{---\texttt{"}---}

\title{Improving noise robust automatic speech recognition \\with single-channel time-domain enhancement network}
\name{Keisuke Kinoshita, Tsubasa Ochiai, Marc Delcroix, Tomohiro Nakatani}
\address{NTT Communication Science Laboratories, NTT Corporation, Kyoto, Japan}%
\begin{document}
\ninept
\maketitle
\begin{abstract}
With the advent of deep learning, research on noise-robust automatic speech recognition (ASR) has progressed rapidly.
However, ASR performance in noisy conditions of single-channel systems remains unsatisfactory.
Indeed, most single-channel speech enhancement (SE) methods (denoising) have brought only limited performance gains over state-of-the-art ASR back-end trained on multi-condition training data.
Recently, there has been much research on neural network-based SE methods working in the time-domain showing levels of performance never attained before. However, it has not been established whether the high enhancement performance achieved by such time-domain approaches could be translated into ASR.
In this paper, we show that a single-channel time-domain denoising approach can significantly improve ASR performance, providing more than 30 \% relative word error reduction over a strong ASR back-end on the real evaluation data of the single-channel track of the CHiME-4  dataset. These positive results demonstrate that single-channel noise reduction can still improve ASR performance, which should open the door to more research in that direction. 
\end{abstract}

\section{Introduction}
\label{sec:introduction}
Recently, the development of deep learning technologies has led to great progress in the field of automatic speech recognition (ASR).
Current state-of-the-art ASR systems are approaching human recognition performance levels~\cite{Xiong2018_Microsoft2017ConversationalSpeech,IBM_switchboard17}, 
when speech is recorded with a close-talking microphone.
However, recognition of speech recorded by distant microphones remains challenging 
because of acoustic interference such as noise, reverberation and interference speakers.

The problem of distant ASR has attracted increasing attention.
When a microphone array is available, ASR performance can be greatly improved by employing multi-channel speech enhancement (SE) pre-processing 
with an ASR back-end trained on multi-condition training (MCT) data.
For example, the combination of neural-network (NN) based time-frequency mask estimation with beamforming has been employed by all top systems in recent distant ASR challenges~\cite{heymann2017beamnnet,CHiME4_results}. 
It is worth mentioning that multi-channel SE can improve ASR performance even without any retraining of the ASR back-end on the enhanced speech, 
which may be possible because they introduce only a few distortions to the processed signals.
However, there are many situations where only a single microphone is available.
In such cases, the ASR performance remains far behind that obtained with a microphone array~\cite{CHiME4_results}.
Therefore, there is a need for more research on effective single-channel SE front-ends for ASR.

There has been much research on single-channel SE for noise reduction and speech separation~\cite{kolbeak2018single}.
Early deep learning-based approaches used NNs that operate in the frequency-domain to predict a time-frequency mask that reduces noise 
when applied to the microphone signal\cite{fu2017raw,chen2017long,wang2014training,erdogan2015recoboost,williamson2016complex}. 
Although these frequency-domain approaches have successfully improved SE evaluation metrics, e.g., signal-to-distortion ratio (SDR), 
this improvement did not lead to better ASR performance.
For example, NN-based masking~\cite{yoshioka2015ntt,chen2018building} and other conventional single-channel enhancement techniques~\cite{fujimoto2019onepass} 
did not contribute to the ASR performance improvement, when used with a state-of-the-art ASR back-end~\cite{chen2018building}. 
This suggests that most single-channel SE approaches tend to introduce distortions that create a mismatch with the ASR back-end, 
therefore limiting their effect on ASR.

Recently, SE operating directly in the time domain has received increased interest.
There has been much research proposing time-domain NNs for noise reduction~\cite{pascual2017segan,rethage2018wavenet,fu2018end} and speech separation~\cite{luo2018tasnet,luo2019conv}.
These approaches have achieved a level of SE performance never attained before~\cite{luo2019conv}.
However, time-domain approaches have not been sufficiently investigated in the context of noise-robust ASR.

In this paper, we investigate the use of time-domain NN for noise reduction to examine whether the great SE performance improvement can be translated into ASR.
Motivated by the success of temporal convolutional NN-based architecture in speech separation~\cite{luo2019conv} and various sequence modeling tasks~\cite{bai2018empirical}, we base the NN architecture for our investigation on the recently proposed convolutional time-domain audio separation network (Conv-TasNet)~\cite{luo2019conv}, which has achieved state-of-the-art performance surpassing even frequency-domain ideal masking.  
We adapt Conv-TasNet for the noise reduction task and call it Denoising-TasNet. A similar network has been used in~\cite{kim2018multidomain}.
We investigate two variants of Denoising-TasNet, one predicting only the enhanced speech and one with two outputs predicting speech and noise.
The latter enables defining a multi-task loss, which can regularize the network training and is shown to achieve better ASR performance.

We perform experiments on CHiME-4 data~\cite{chime4}, which includes real noisy recordings that are particularly relevant to our investigation. 
The findings of this paper are that 
(1) Denoising-TasNet does not only achieve high SE performance but also significantly reduces ASR word error rates (WERs) on real recordings even when used with a strong ASR back-end trained on MCT data ( more than 30\% relative WER reduction);  
(2) it can improve ASR performance even without retraining the ASR back-end; 
(3) interestingly, by augmenting the amount of training data for the SE front-end, we could improve the performance of Denoising TasNet, but also of a frequency-domain BLSTM-based masking approach;
(4) Finally, we show that the performance improvement can be generalized to a different task, i.e. AURORA-4~\cite{Aurora4}.  
These positive results demonstrate that single-channel noise reduction can still improve ASR performance, which should open the door to more research in that direction.

\section{Denoising network}
In this section, we first introduce the notations and then describe the frequency-domain and time-domain denoising networks that we use in our experiments. 
Finally, we discuss the loss functions used to train the networks.

\begin{figure}[t]
  \centering
  \centerline{\includegraphics[width=7cm]{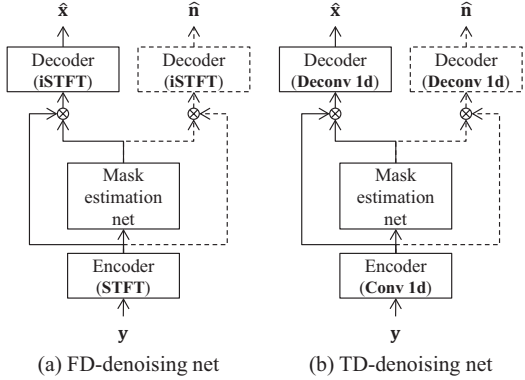}}
\vspace{-2mm}
\caption{Block diagrams of denoising NNs investigated in this study. Parts of the systems shown with dashed line are optional network branches to perform multi-task training mentioned in \ref{sec:loss}.}
\label{fig:diagram}
\end{figure}

\subsection{Problem formulation}
Let us consider a single-channel microphone signal, $y(t)$,
\begin{equation}
    y(t) = x(t) + n(t),
\end{equation}
where $x(t)$ is the target speech signal, $n(t)$ is background noise, and $t$ is a time index. 
In this paper, the background noise does not include obvious interfering speakers.
We define the vector representation of the time-domain signals as $\mathbf{y}=[y(1), \ldots, y(T)]$, where $T$ is the length of the microphone signal. We aim at recovering the speech signal $\mathbf{x}$ from the microphone signal $\mathbf{y}$.

\subsection{Time-domain and frequency-domain denoising networks}
Figure \ref{fig:diagram} is a schematic diagram of the denoising NN framework that we investigate in (a) the frequency-domain and (b) the time-domain.
The denoising process is performed in 3 steps, (1) encoding, (2) mask estimation, (3) decoding.

{\bf (1) Encoding:}
First, in the encoding step, we obtain an encoded version of the microphone signal, $\mathbf{e}_y$, using an encoder module as,
\begin{equation}
    \mathbf{e}_y = \text{encoder}(\mathbf{y}),
\end{equation}
where $\text{encoder}(\cdot)$ represents the encoder function. 
For frequency-domain networks, the encoder function consists of the short-time Fourier transform (STFT).
For time-domain network, the encoder function consists of a learnable 1D convolution layer, 
followed by a ReLu activation function.
The encoded domain learned through the NN training may be less interpretable compared to STFT, 
but allows a signal representation optimized for denoising.

{\bf (2) Mask estimation: }
Second, we estimate a denoising mask using a mask estimation network. 
The mask estimation NN should account for the time context of the signal to distinguish speech from noise. This can be achieved by using BLSTM layers or dilated-convolution layers as in TasNet. The output of the mask estimation network can be either a single mask used to predict speech or two masks, one for speech and another for noise prediction as,
\begin{equation}
[ \mathbf{m}_x,\mathbf{m}_n ] = \text{MEnet}(\mathbf{e}_y),
\end{equation}
where $\text{MEnet}(\cdot)$ is a mask estimation network, and $\mathbf{m}_x$ and $\mathbf{m}_n$ are the masks associated with speech and noise, respectively.

When working in the frequency-domain, we usually compute the amplitude of the STFT coefficients to work on real numbers before inputting them to the mask estimation network.  
In the case of frequency-domain denoising, the masks are conventional time-frequency soft masks that indicate to what degree the target speech is dominant.
For time-domain networks, the masks are defined in the encoded domain of the encoder.

{\bf (3) Decoding: }
Finally, we obtain the enhanced speech signal $\hat{\mathbf{x}}$ by applying the speech mask, $\mathbf{m}_x$ to the encoded microphone signal, $\mathbf{e}_y$, and applying a decoding function as,
\begin{equation}
\hat{\mathbf{x}} = \text{decoder}(\mathbf{m}_x \odot \mathbf{e}_y),
\end{equation}
where $\text{decoder}(\cdot)$ is the decoder function, $\odot$ is an element-wise multiplication.
For frequency-domain denoising, the decoder consists of inverse STFT.
For time-domain, it consists of a learnable 1D deconvolution layer.

Note that when the mask estimation network also outputs a noise mask, it is possible to obtain an estimate of the noise signal as,
\begin{equation}
\hat{\mathbf{n}} = \text{decoder}(\mathbf{m}_n \odot \mathbf{e}_y).
\end{equation}
In this paper, the estimated noise is used only during training to derive a multi-task training loss as described below.

\subsection{Training losses}
\label{sec:loss}

\qquad {\bf Time-domain loss (TDL): }
For time-domain networks, we employ the classic signal-to-noise ratio (SNR)~\cite{roux2019sdr} as time-domain loss,
\begin{equation}
    L(\theta) = -\text{SNR}(\mathbf{x}, \hat{\mathbf{x}}) ,
\end{equation}
where $\theta$ are the model parameters, $\text{SNR}=-10 \log_{10}(\frac{||\mathbf{x}||^2}{||\mathbf{x}-\hat{\mathbf{x}}||^2})$ is the SNR between the clean speech and the enhanced speech, and $||\cdot||^2$ is the $L^2$ norm. 
We decided to use the classic SNR loss~\cite{roux2019sdr} instead of the scale-invariant SNR (SiSNR) used in the original TasNet~\cite{luo2018tasnet}, because training the network with SiSNR let the network freely change the level of the enhanced signal.
With the SNR loss, the scale of the signals is preserved avoiding any scaling requirement when passing the signal to ASR.
Moreover, we confirmed in preliminary experiments that using the SNR instead of SiSNR lead to slightly better SE performance.

We can also train the frequency-domain denoising networks with a TDL 
by computing the loss on the reconstructed time-domain signals, i.e. after the inverse STFT decoder.

{\bf Frequency-domain loss (FDL): }
For frequency-domain networks, we use the log mean square error loss (log-MSE) as frequency-domain loss,
\begin{equation}
    L(\theta) = - 10 \log_{10}(|| \ |\mathbf{e}_x|-|\mathbf{m}_x \odot \mathbf{e}_y| \ ||^2),
\end{equation}
where $|\mathbf{e}_x|$ is the amplitude spectrum of the target speech signal.
The FDL is computed before the decoder (iSTFT). 
Note that the log-MSE is equivalent to the SNR loss computed on the amplitude spectrum.

{\bf Noise reconstruction loss (NR-loss): }
In addition to the above losses, 
we investigate using a multi-task loss by adding a second noise reconstruction loss similar to~\cite{Erdogan2018}. 
For example, in the case of the time-domain loss, the multi-task loss is defined as, 
\begin{equation}
    L(\theta) = -(\text{SNR}(\mathbf{x}, \hat{\mathbf{x}}) + \text{SNR}(\mathbf{n}, \hat{\mathbf{n}})).
\end{equation}
Forcing the network to predict the noise signal may act as a regularizer enabling to train more robust models.

Note that TasNet was originally proposed for speech separation, 
and thus requires permutation invariant training (PIT)~\cite{kolbaek2017multitalker} to solve a permutation ambiguity of the output speech sources.
Here, since the NN output is speech and noise, we can fix the output order of the network and avoid using permutation invariant training~\cite{kolbaek2017multitalker}.

\section{Related works}

There have been several works combining single-channel frequency-domain SE for noise-robust ASR.
GAN-based training for SE~\cite{pascual2017segan} has received increased attention.
GAN is employed to constrain the estimated signals close to the clean signals, which was shown to 
improve objective and subjective SE criterion, 
but it did not contribute to improvement in terms of ASR~\cite{donahue2018exploring}.

Several recent works have also tackled the single-channel CHiME-4 task.
In~\cite{fujimoto2019onepass}, several legacy noise reduction approaches were evaluated for online ASR. 
Although the SE front-ends did not improve ASR when used individually, about 20 \% relative WER reduction was possible when used in multi-branch network architecture  (from 22.6 \% to 19.4 \% WER  on CHiME-4 eval real without and with SE front-end, respectively). However, this study used an online ASR system, which resulted in a much worse baseline than ours.

Another work proposed to use a parametric Wiener Filter~\cite{Lim_IEEE79} with a mask estimation network as front-end and reported performance gains with a strong ASR back-end when the hyper-parameters of the Wiener filter were optimized based on the acoustic-level criterion~\cite{menne2019investigation}.
They achieved about 8\% relative WER reduction (from 11.6 \% to 10.6 \% WER  on CHiME-4 eval real without and with SE front-end, respectively).

Compared to these previous studies, we focus on comparing frequency-domain and time-domain SE.
Tighter integration of time-domain SE front-end with the ASR back-end such as~\cite{menne2019investigation,fujimoto2019onepass} is one of our future research directions.

\section{Experiments}
\label{sec:exp}
We perform extensive experiments on CHiME-4 data to evaluate the performance difference between frequency domain and time domain approaches, and the effect of the loss functions, i.e. TDL, FDL and NR-loss.
We evaluated the performance in terms of SDR~\cite{vincent2006performance} and WER.
All experiments were performed on speech sampled at 16kHz. 

For evaluation, we use the official evaluation set of the CHiME-4 corpus, which consists of a noisy version of WSJ0 utterances recorded in 4 environments, street, cafe, bus, and pedestrian at average SNR of about 5 dB~\cite{barker2015third}.
The corpus consists of simulated and real noisy recordings based on a microphone array attached to a tablet.
The test set includes 1320 simulated and 1320 real recordings.

\subsection{SE systems: network configurations and training data}
In this experiment, we compared the following 3 different network configurations.

{\bf Denoising-TasNet: }
We investigated the performance of the Denoising-TasNet that uses a similar configuration to the original Conv-TasNet~\cite{luo2019conv}.
Our implementation is based on the open-source implementation of Conv-TasNet~\cite{funcwj}.
In particular, following the hyper-parameter notations in the original paper~\cite{luo2019conv}, 
we set the hyper-parameters to N=256, L=20, B=256, H=512, P=3, X=8, R=4. We used Adam optimizer~\cite{Kingma2014Adam} to train the network with a learning rate of 1e-3.

{\bf Frequency-domain BLSTM network: }
We compared Denoising-TasNet with a frequency-domain BLSTM (FD-BLSTM) network, 
which uses a mask estimation network consisting of 3 BLSTM layers with 896 units followed by a linear layer with ReLU activation~\cite{kolbaek2017multitalker}.
The input of the mask estimation network consists of amplitude spectrum coefficients computed with an STFT with a hanning window of 32 msec and a shift of 8 msec. 
We trained two versions of the FD-BLSTM denoising network, one with FD loss (FD-BLSTM-FDL) and one with TD loss (FD-BLSTM-TDL). 
We based our implementation on~\cite{funcwj}. We set the learning rate of Adam to 1e-3.

{\bf Frequency-domain Convolutional network: }
We also compared Denoising-TasNet with another frequency-domain network that replaces the BLSTM-based mask estimation network with a
convolutional architecture similar to that of the Denoising-TasNet. We employed the same training strategy as for Denoising-TasNet.
We trained an FD-Conv denoising network with FD loss (FD-Conv-FDL).

Our preliminary experiments confirmed that 
using the official training set of the CHiME-4 corpus to train the SE systems always leads to poorer results.
This is probably because it does not contain reverberation~\cite{barker2015third}.
Thus, for the training of the SE systems, we created an alternative noisy and reverberant training set based on the image method~\cite{allen1979image}.
For each utterance, we randomly generated a simulated room impulse responses (RIRs) 
and selected a noise signal from the CHiME-4's official noise recordings.
Here, we set the T60s between 0.2 and 0.7~s, and the speaker-microphone distance between 10 and 60~cm randomly.
Based on the RIRs and noises, we created the same number of utterances as the official training set,
i.e., the 35690 simulated training utterances (referred to as ``35k set'' hereafter), at SNR randomly selected between 0 and 5~dB\footnote{Our setting was found to slightly diverge from the CHiME-4 challenge regulation because the challenge did not allow random selection of noise samples and SNRs
for training data generation.}.

\subsection{ASR back-end configurations and training data}
We used Kaldi to build a hybrid DNN-HMM ASR back-end.
It consists of a time-delayed NN-based AM trained with the lattice-free MMI (LF-MMI) objective function.
We use a forward-backward LSTM-LM for language model rescoring. 
This back-end follows the standard Kaldi recipe.

We trained two types of acoustic models (AMs) that differ in training data;
One AM, referred to as Org-AM, is 
trained on data including the aforementioned 35k set 
and the official training data of CHiME-4 comprising 52428 utterances.
Another AM, referred to as Enh-AM, was trained on data
including the aforementioned 35k set, its enhanced version 
and the official training data of CHiME4.
Enh-AMs are created to simulate the effect of retraining on enhanced signals.

\subsection{Results}

\subsubsection{Effect of NR-loss}

\begin{table}[t]
  \renewcommand{\arraystretch}{0.95}
  \caption{Effect of NR-loss on two representative SE front-ends.}
\vspace{1mm}
  \label{tab:res_mtl}
  \centering
  \scalebox{0.9}{
  \begin{tabular}{ l c c c | c }
    \toprule
                    & NR-loss     & \multicolumn{2}{c|}{WER} &  SDR \\
    System          &         & Simu   & Real           & Simu  \\
    \midrule
    FD-BLSTM-FDL    & ---        & 15.13   & 13.91           & 11.84  \\
                    &\checkmark  & 13.89   & 12.89           & 11.96  \\
    \midrule
    Denoising-TasNet &---        & 12.31   & 10.64   & 14.13 \\
                     &\checkmark & 11.87   & 9.75   & 14.21  \\           
    \bottomrule
  \end{tabular}
  }
\vspace{-5mm}
\end{table}

We first examine the effect of NR-loss described in \ref{sec:loss} 
to determine if it is beneficial for ASR.
Table \ref{tab:res_mtl} shows the WERs with and without NR-loss. 
Interestingly, we observed that although using NR-loss did not improve SDR significantly,
it consistently led to improved ASR performance.
Here we show the results of only two representative methods, 
but we observed the same tendency for all the SE front-end variants we tested.
Therefore, we use NR-loss in all following experiments.

\subsubsection{Effect of different SE systems}
Table~\ref{tab:res_CHiME-4} shows the results for the four SE front-end systems investigated, with Org-AM and Enh-AM.
``No process'' in the table refers to the results obtained with unprocessed noisy observation.

The table shows that FD-BLSTM-FDL improves SDR but not WER compared to the baseline system (``No process'').
This result is consistent with the finding of the conventional studies~\cite{yoshioka2015ntt,chen2018building}.
However, in case the acoustic model knows the artifacts that the SE front-end may induce, i.e., the Enh-AM case, we could achieve a moderate relative WER reduction of up to 10 \%.

FD-BLSTM-TDL and FD-Conv-FDL succeeded to further improve the SDR by utilizing TDL and convolution-based architecture, respectively.
However, the ASR performance dropped significantly.
We hypothesize that although more noise could be removed, these systems introduced more distortions that reduced ASR performance even for the Enh-AM case.

On the other hand, Denoising-TasNet achieved not only the highest SDR but also improved ASR performance significantly.
Interestingly, Denoising-TasNet improved ASR even without retraining the back-end, i.e., the Org-AM case.
With the Enh-AM, we achieved more than 30\% relative error rate reduction on the real recordings, 
which is a significantly higher improvement than conventional FD-BLSTM-based approaches.

\begin{table}[t]
  \renewcommand{\arraystretch}{0.95}
  \caption{SDR [dB] and WER [\%] for CHiME-4 corpus}
\vspace{1mm}
  \label{tab:res_CHiME-4}
  \centering
  \scalebox{0.9}{
  \begin{tabular}{ l  c c c | c }
    \toprule
    &   AM  &\multicolumn{2}{c|}{WER} & SDR  \\
    System  &   &   Simu       & Real      & Simu \\
    \midrule
    No process & Org-AM     & 12.53      & 12.23     & 5.09 \\ 
    \midrule
    FD-BLSTM-FDL    & Org-AM     & 13.89     & 12.89      & 11.96 \\
                         & Enh-AM     & 11.66     & 11.11      & \ditto \\
    \midrule
    FD-BLSTM-TDL    & Org-AM     & 18.46      & 18.34     & 13.52 \\
                          & Enh-AM     & 13.39      &13.28     & \ditto\\ 
    \midrule
    FD-Conv-FDL & Org-AM      & 22.96     & 24.32      & 12.74  \\
                         & Enh-AM      & 15.91     & 16.48      & \ditto   \\
    \midrule
    Denoising-TasNet    & Org-AM      & 11.87     & 9.75      & 14.21  \\
                         & Enh-AM      & 9.88     & 8.19       & \ditto \\
    \bottomrule
  \end{tabular}
   }
\vspace{-3mm}
\end{table}

\begin{table}[t]
   \renewcommand{\arraystretch}{0.95}
   \caption{Effect of data augmentation for CHiME-4 corpus.}
\vspace{1mm}
\label{tab:res_adagaug}
  \centering
  \scalebox{0.9}{
  \begin{tabular}{ l c c c | c }
    \toprule
            &           AM & \multicolumn{2}{c|}{WER} & SDR  \\
    System                &            & Simu       & Real      & Simu \\
    \midrule
    FD-BLSTM-FDL                         & Org-AM      & 12.25      & 10.50    & 11.87 \\
                                         & Enh-AM      & 10.40      & 8.88     & \ditto \\
    \midrule                
    Denoising-TasNet                     & Org-AM      & 10.82      & 8.33     & 14.24 \\
                                         & Enh-AM      & 9.67       & 7.68     & \ditto \\
     \bottomrule
   \end{tabular}
   }
\vspace{-3mm}
\end{table}

\subsubsection{Effect of data augmentation on SE systems}
We augmented the variation of the noisy and reverberant training data to create up to 100000 utterances (100k set) and trained the SE systems on this 100k set to see the effect of data augmentation. 

Table~\ref{tab:res_adagaug} shows the WERs and SDRs of the CHiME-4 test set obtained with two representative SE systems trained on the 100k set.
For this experiment, we used the same ASR backend configurations as in Table~\ref{tab:res_CHiME-4}.
Augmenting training data from 35k to 100k for the SE systems appears to significantly improve the ASR performance, 
while it has only a little impact on SDRs.
Interestingly, although these results still support the superiority of Denoising-TasNet over FD-BLSTM,
they also show the great potential of single-channel SE systems themselves, 
since even FD-BLSTM could greatly improve ASR performance when trained on the augmented dataset.
Note that we found that adding the same 100k set to the training data for AMs did not yield any improvement in the baseline ASR performance.

\subsubsection{Generalization to a different task}
We examine the generalization capability of the SE systems to different noise conditions and different ASR back-ends.
For this experiment, we used the Aurora-4 dataset, which also consists of a noisy version of WSJ with different types and levels of noise than CHiME-4.
The noise types include street traffic, train terminals, and stations, cars, babble, restaurants, and airports. 
SNRs for the test data ranges from 5 to 15 dB. 
As we focus on denoising, we only report the results of the test sets A (clean speech) and B (noisy speech).
For these experiments, we use a CTC-attention hybrid end-to-end ASR back-end from the open-source toolkit~\cite{espnet,espnet_web} to build the ASR back-end. 
The details of the configuration can be found in~\cite{espnet_aurora4}. 

Table \ref{tab:res_Aurora-4} shows the results of FD-BLSTM and Denoising-TasNet on the Aurora-4 dataset. We used the SE systems trained with the 100k set. We confirmed that both SE systems could successfully generalize to this task, and the superiority of denoising TasNet is still maintained. Note that the degradation in set-A by denoising TasNet is not significant.

\begin{table}[t]
  \renewcommand{\arraystretch}{0.95}
  \caption{WER [\%] for Aurora-4 corpus. }
\vspace{1mm}
  \label{tab:res_Aurora-4}
  \centering
  \scalebox{0.9}{
  \begin{tabular}{ l c c }
    \toprule
    System & set-A (clean) & set-B (noisy)\\
    \midrule
    No process & 4.3 & 8.5 \\
    \midrule
    FD-BLSTM-FDL     &  4.2  & 7.7  \\
    Denoising-TasNet  & 4.4 & 6.3 \\
    \bottomrule
  \end{tabular}
  }
\vspace{-3mm}
\end{table}

\section{Conclusion}
\label{sec:conclusion}

We have experimented different configurations of frequency-domain and time-domain denoising NNs.
We observed that Denoising-TasNet significantly improved ASR performance by more than 30\% on real recordings 
with a strong ASR back-end.
Interestingly, we also found that 
not only Denoising TasNet but also FD-BLSTM could significantly improve the ASR performance,
when these SE systems were trained with a large training dataset, i.e., 100k set.
Such a large improvement demonstrates that single-channel denoising still has the potential for improving the noise robustness of ASR systems.

\footnotesize
\bibliographystyle{IEEEbib}
\bibliography{strings,refs}
\end{document}